\documentclass[12pt,letterpaper]{JHEP3}
\usepackage{cite}
\usepackage{epsfig}

\title{TASI Lectures on the Cosmological Constant}

\author{%
Raphael Bousso\\ 
Center for Theoretical Physics, Department of Physics\\
University of California, Berkeley, CA 94720-7300, U.S.A.\\
{\em and}\\
Lawrence Berkeley National Laboratory, 
Berkeley, CA 94720-8162, U.S.A.}

\abstract{%
  The energy density of the vacuum, $\Lambda$, is at least 60 orders
  of magnitude smaller than several known contributions to it.
  Approaches to this problem are tightly constrained by data ranging
  from elementary observations to precision experiments. Absent
  overwhelming evidence to the contrary, dark energy can only be
  interpreted as vacuum energy, so the venerable assumption that
  $\Lambda=0$ conflicts with observation.  The possibility remains
  that $\Lambda$ is fundamentally variable, though constant over large
  spacetime regions.  This can explain the observed value, but only in
  a theory satisfying a number of restrictive kinematic and dynamical
  conditions.  String theory offers a concrete realization through its
  landscape of metastable vacua.
}

\begin{document}

\section{Introduction: The cosmological constant problem}
\label{sec-ccp}

When Einstein wrote down the field equation for general relativity, 
\begin{equation}
  R_{\mu\nu} - \frac{1}{2} R g_{\mu\nu} + \Lambda g_{\mu\nu} =
  8\pi G T_{\mu\nu}
\end{equation}
he had a choice: The cosmological constant $\Lambda$ was not fixed by
the structure of the theory.  There was no formal reason to set it to
zero, and in fact, Einstein famously tuned it to yield a static
cosmological solution---his ``greatest blunder''.

The universe has turned out not to be static, and $\Lambda$ was
henceforth assumed to vanish.  This was never particularly satisfying
even from a classical perspective.  The situation is similar to a
famous shortcoming of Newtonian gravity: Nothing prevents us from
equating the gravitational charge with inertial mass, but nothing
forces us to do so, either.

Any nonzero value of $\Lambda$ introduces a length scale and time
scale
\begin{equation}
r_\Lambda= c t_\Lambda=\sqrt{3/|\Lambda|}
\label{eq-rlam}
\end{equation}
into Einstein's theory.  An independent, natural length scale arises
from the constants of nature: the Planck length\footnote{Here $G$
  denotes Newton's constant and $c$ is the speed of light. In this
  paper Planck units are used unless other units are given explicitly.
  For example, $t_{\rm P}=l_{\rm P} /c \approx .539 \times 10^{-43}
  {\rm s}$ and $M_{\rm P} = 2.177\times 10^{-5} {\rm g}$.}
\begin{equation}
  l_{\rm P} = \sqrt{\frac{G\hbar}{c^3}}
  \approx 1.616 \times 10^{-33} {\rm cm}~.
\end{equation}

Whether $|\Lambda|$ vanishes or not, it has long been known
empirically that it is very small in Planck units (i.e., that
$r_\Lambda$ is large in these natural units).  The cosmological
constant strongly affects spacetime dynamics at all scales larger than
$r_\Lambda$ and $t_\Lambda$.  But we see general relativity operate on
scales much larger than the Planck length, without any sign of the
cosmological constant.  In fact, the smallness of $\Lambda$ can be
deduced just from the fact that the universe is large compared to the
Planck length, and old compared to the Planck time.

First, consider the case of positive $\Lambda$.  Assume, for the sake
of argument, that no matter is present ($T_{\mu\nu}=0$).  Then the
only isotropic solution to Einstein's equation is de~Sitter space,
which exhibits a cosmological horizon of radius
$r_\Lambda$~\cite{HawEll}.  A cosmological horizon is the largest
observable distance scale, and the presence of matter will only
decrease the horizon radius~\cite{GibHaw77a}.  We see scales that are
large in Planck units ($r\gg 1$), and since $r_\Lambda$ must be even
larger, Eq.~(\ref{eq-rlam}) implies that the cosmological constant is
small.

Negative $\Lambda$ causes the universe to recollapse independently of
spatial curvature, on a timescale $t_{\Lambda}$~\cite{Edw72}.  The
obvious fact that the universe is old compared to the Planck time then
implies that $(-\Lambda)$ is small.  Summarizing the above arguments,
we find
\begin{equation}
-3t^{-2}\lesssim\Lambda\lesssim 3r^{-2}~,
\label{eq-trlam}
\end{equation}
where $t$ and $r$ are any time scale and any distance scale that have
been observed.

These conclusions did not require cutting-edge experiments: knowing
only that the world is older than 5000 years and larger than Belgium
would suffice to tell us that $|\Lambda|\ll 1$.  For a tighter
constraint, note that we can see out to distances of billions of light
years, so $r>10^{60}$; and stars are billions of years old, so
$t>10^{60}$.  With these data, known for many decades,
Eq.~(\ref{eq-trlam}) implies roughly that
\begin{equation}
  |\Lambda|\lesssim 3\times 10^{-120}~.
\label{eq-small}
\end{equation}
Hence $\Lambda$ is very small indeed.

This result makes it tempting to set $\Lambda=0$ in the Einstein
equation and move on.  But $\Lambda$ returns through the back door.
The quantum fluctuations in the vacuum of the standard model
contribute to the expectation value of the stress tensor in a way that
mimics a cosmological constant. It is this effect that turns the
cosmological constant from a mere ambiguity into a genuine problem.

In quantum field theory, the vacuum is highly nontrivial.  As a
harmonic oscillator in the ground state, every mode of every field
contributes a zero point energy to the energy density of the vacuum.
In a path integral description, this energy arises from virtual
particle-antiparticle pairs, or ``loops'' (Fig.~\ref{fig-loop}a).  By
Lorentz invariance, the corresponding energy-momentum-stress tensor
had better be proportional to the metric,
\begin{equation}
\langle T_{\mu\nu} \rangle = -\rho_{\rm vacuum} g_{\mu\nu} ~,
\label{eq-lrho}
\end{equation}
which is confirmed by direct calculation.  

Though it appears on the right hand side of Einstein's equation,
vacuum energy has the form of a cosmological constant, and we might as
well absorb it and redefine $\Lambda$ via
\begin{equation}
  \Lambda=\Lambda_{\rm Einstein}   +8\pi\rho_{\rm vacuum}~.
\end{equation}
Equivalently, we may absorb the ``bare'' cosmological constant
appearing in Einstein's equation, $\Lambda_{\rm Einstein}$, into the energy
density of the vacuum, defining
\begin{equation}
\rho_\Lambda\equiv \rho_{\rm vacuum}+\frac{\Lambda_{\rm Einstein}}{8\pi}~.
\end{equation}
\EPSFIGURE{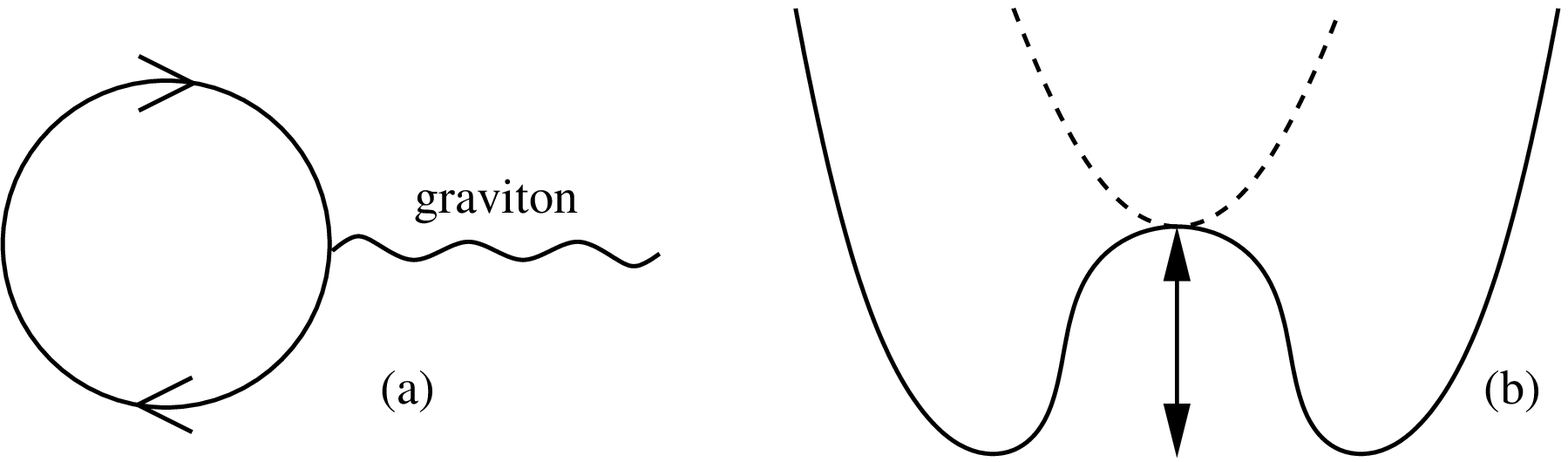,width=.8\textwidth}{\label{fig-loop} Some
  contributions to vacuum energy.  (a) Virtual particle-antiparticle
  pairs (loops) gravitate.  The vacuum of the standard model abounds
  with such pairs and hence should gravitate enormously. (b) Symmetry
  breaking in the early universe (e.g., of the chiral and electroweak
  symmetries) shifts the vacuum energy by amounts dozens of orders of
  magnitude larger than the observed value.}
Eqs.~(\ref{eq-rlam}), (\ref{eq-trlam}), and (\ref{eq-small}) apply to
the total cosmological constant, and can be restated as an empirical
bound on the total energy density of the vacuum:
\begin{equation}
  |\rho_\Lambda|\lesssim 10^{-121}~.
\end{equation}

But in the standard model, the energy of the vacuum receives many
contributions much larger than this bound.  Their value depends on the
energy scale up to which we trust the theory but is enormous even with
a conservative cutoff.

For example, consider the electron, which is well understood at least
up to energies of order $M=$ 100 GeV~\cite{Pol06}.  Dimensional
analysis implies that electron loops up to this cutoff contribute of
order $(100$ GeV$)^4$ to the vacuum energy, or $10^{-68}$ in Planck
units.  Similar contributions are expected from other fields.  The
real cutoff is probably of order the supersymmetry breaking scale,
giving at least $(1$ TeV$)^4\approx 10^{-64}$.  It may be as high as
the Planck scale, which would yield $|\rho_\Lambda|$ of order unity.
Thus, quantum field theory predicts $|\rho_\Lambda|$ to be some 60 to
120 orders of magnitude larger than the experimental bound,
Eq.~(\ref{eq-small}).

Additional contributions come from the potentials of scalar fields,
such as the potential giving rise to symmetry breaking in the
electroweak theory (Fig.~\ref{fig-loop}b).  The vacuum energy of the
symmetric and the broken phase differ by approximately $(200$
GeV$)^4\approx 10^{-67}$.  Any other symmetry breaking mechanisms at
higher or lower energy, such as chiral symmetry breaking of QCD with
$(300$ MeV$)^4\approx 10^{-79}$, will also contribute.

I have exhibited various known contributions to the vacuum energy.
They are uncorrelated with one another and with the (unknown) bare
cosmological constant appearing in Einstein's equation, $\Lambda_{\rm
  Einstein}$.  Each contribution is dozens of orders of magnitude
larger than the empirical bound today, Eq.~(\ref{eq-small}).  In
particular, the radiative correction terms from quantum fields are
expected to be at least of order $10^{-64}$.  They come with different
signs, but it would seem overwhelmingly unlikely for such large terms
to cancel to better than a part in $10^{120}$, in the present era.

This is the cosmological constant problem: {\em Why is the vacuum
  energy today so small?\/}\footnote{Recent observations have revealed
  the actual value of the cosmological constant: $\rho_\Lambda\approx
  1.5\times 10^{-123}$.  This sharpens the issue, but it does not
  change the fact that a much larger value is predicted.  I will focus
  first on the question of why $\rho_\Lambda $ is not large, before
  considering the implications of its precise value
  (Sec.~\ref{sec-duck}).}  It represents a serious crisis in physics:
a discrepancy between theory and experiment, of 60 to 120 orders of
magnitude, in a quantity as basic as the weight of empty space.

\paragraph{Outline} 
In Sec.~\ref{sec-doa}-\ref{sec-notyetdead}, I survey a number of
general approaches to the cosmological constant problem, without going
into detailed theoretical models.\footnote{I will not attempt a
  thorough survey of the literature.  A classic review is
  Ref.~\cite{Wei89}.  Ref.~\cite{Car00} is comprehensive and includes
  the discovery of nonzero $\rho_\Lambda$.  For a particularly clear
  recent discussion that includes the string landscape, see
  Ref.~\cite{Pol06}.  This article is based on lectures at the
  Theoretical Advanced Study Institute, at the University of Colorado,
  Boulder, in May/June of 2007, and is being published under the title
  ``The Cosmological Constant'' in a special issue on dark energy in
  the journal {\em General Relativity and Gravitation}; some material
  has appeared earlier in Ref.~\cite{Bou06b}.}
Sections~\ref{sec-duck} and \ref{sec-theory} discuss recent
experimental and theoretical progress, and Sec.~\ref{sec-outlook}
closes with an outlook.

Many tempting ideas can be ruled out quite generally, because they
conflict with well-tested physics.  In Sec.~\ref{sec-doa}, I discuss a
number of examples.  Their failure modes illustrate the difficulty of
the problem and constitute useful litmus tests for new approaches.  I
emphasize the physical origin of basic obstructions, rather than the
various technical symptoms through which they manifest themselves in
specific models.

In Sec.~\ref{sec-nowdead}, I discuss an example of a class of ideas
that predict that $\rho_\Lambda=0$ today.  This strategy did seem
viable once, but it never found a concrete theoretical realization,
and it is now quite disfavored experimentally.

In Sec.~\ref{sec-notyetdead}, I discuss the idea that $\rho_\Lambda$
is a dynamical variable that can take on different values in
different, large parts of the universe.  One can show that galaxies
form only in regions where $\rho_\Lambda$ is not very much larger than
the present matter density.  Moreover, it is reasonable to suppose
that regions without galaxies do not contain any observers.  This
combination of arguments, first formulated by Weinberg in 1987,
predicts that we should observe $\rho_\Lambda \sim \rho_{\rm matter}$.
The approach makes a number of theoretical predictions as well, namely
that its stringent kinematic and dynamical requirements could
eventually be accommodated in a realistic theory.

In Sec.~\ref{sec-duck} I argue that cosmological observations since
1998 have discriminated powerfully between approaches to the
cosmological constant problem.  Weinberg's prediction was borne out by
the observation of phenomena such as the accelerated expansion and
spatial flatness of the universe, which are incompatible with
$\rho_\Lambda=0$ and favor a value $\rho_\Lambda\approx 3.2\,\rho_{\rm
  matter}$.

Theory, too, has come down on Weinberg's side.  In
Sec.~\ref{sec-theory}, I describe a theory of the cosmological
constant.  It satisfies the theoretical requirements of his approach,
and it arises naturally in string theory.

String theory was not invented for the purpose, and it is a rather
rigid framework.  It is all the more remarkable that some of its most
characteristic features, such as extra dimensions and branes, have
allowed it to address the cosmological constant problem.  They give
string theory the vacuum structure of a multidimensional potential
landscape.  What we call ``the'' vacuum is but one of perhaps
$10^{500}$ long-lived, metastable vacua, arising from the
combinatorics of branes wrapping handles in compact, extra dimensions.
Our ``big bang'' was actually the decay of a more energetic vacuum.
And our own vacuum, too, will decay.

The landscape of string theory opens new fields of inquiry, a few of
which I sketch in Sec.~\ref{sec-outlook}.  Vacuum energy aside, field
content, masses, and couplings are all expected to vary from vacuum to
vacuum.  This raises questions of predictivity.  The existence of a
multitude of metastable solutions is an essential feature of any
theory describing our world, such as the standard model.  Effective or
statistical descriptions have proven very powerful, and we must learn
to develop analogous methods for string theory.  This is complicated
by the fact that in a gravitational theory, false vacua grow faster
than they decay, producing infinite volumes.  A cosmological measure
or cutoff is needed in order to compute relative probabilities of
different vacua.

\section{Some ideas, and why they don't work}
\label{sec-doa}

A discrepancy by a factor of $10^{60}$ is impressive but, given no
other information, might be shrugged off as just another hierarchy
problem.  However, the cosmological constant problem is far more
severe.  In order to appreciate its unique features and extraordinary
difficulty, it is instructive to consider a few approaches that might
come to mind, and to exhibit some of the obstructions they face.  (My
emphasis on the physical origin of these obstructions, rather than
their technical manifestations, is inspired by a similar viewpoint
compellingly advocated in Polchinski's review~\cite{Pol06}.)

In this and the following sections, I will only consider experimental
constraints that have long been known.  I will not use results from
recent precision cosmological experiments, such as the 1998 discovery
of accelerated expansion.  Their impact will be considered in
Sec.~\ref{sec-duck}.

\subsection{Quantum gravity} 

Fundamentally, the problem amounts to a clash between particle
physics, which sources vacuum energy through quantum effects, and
gravity, which responds to it classically.  To describe this situation
accurately, perhaps we need quantum gravity, which we do not
understand well enough.  So we cannot trust the above arguments.

But it is not so easy to sweep the cosmological constant problem under
the rug of our ignorance.  All matter is quantum mechanical.  Yet, its
large scale gravitational interactions are accurately described by
feeding the expectation value of the stress tensor into classical
general relativity.  A quantum theory of gravity, like any other
extension of our theoretical framework, may help with the cosmological
constant problem---string theory certainly does---but it cannot do so
merely by failing to reproduce semi-classical gravity in the
appropriate limit.

Quantum gravity would be needed to describe loop momenta exceeding
$10^{19}$ GeV, or curvature radii smaller than $10^{-33}$ cm.  To
exhibit the cosmological constant problem it is not necessary to
appeal to such extreme regimes; it arises well inside the regime of
validity of both gravity and quantum field theory.  For example, by
Eq.~(\ref{eq-rlam}), electron loops with momenta up to just 1 MeV
alone should curl up the universe to about a million kilometers---a
small world, but well described by general relativity.

\subsection{Infrared or ultraviolet modifications of gravity} 

This last example makes it clear that classical modifications of
gravity are of no apparent use either.  We can only modify the theory
on scales where it has not been tested.  But the above example falls
into a regime where gravity is well constrained experimentally.

More generally, short-distance modifications are irrelevant since the
smallness of the cosmological constant manifests itself through the
large size of the universe.  The universe is much larger than the
smallest distance scale at which gravity has been tested (fractions of
a millimeter).  At intermediate scales, we can trust general
relativity, and we know that it would have responded to the large
vacuum energy predicted by the standard model.  So we can be sure that
the vacuum energy is really unnaturally small, or zero.

Long-distance (infrared) modifications are unhelpful because we know
that the universe started out small, and the cosmological constant
problem is the prediction that the horizon should have never become
larger than Planck size generically, or at most 100 $\mu$m in some
models.  This would have been a true event horizon, so causality would
have prevented larger scales from playing any dynamical role.  In
particular, a modification of gravity on the present horizon scale
would never have come into play.\footnote{This problem affects any
  approach in which the present horizon size, or some other large
  scale, appears directly as input.  This is inevitably circular,
  since the smallness of the cosmological constant is a necessary
  condition for the largeness of the universe.  If we start by
  assuming its largeness, there is nothing left to explain.}

In summary, the shortest and longest distances are precisely the ones
that play no essential role in the cosmological constant problem.  We
know that vacuum energy is unnaturally small ($\rho_\Lambda\ll 10^{-60}$),
assuming only that general relativity is valid on at least one
intermediate scale between 100 $\mu$m and 1 Gpc.  This includes broad
regimes where we know that general relativity is very accurate and
experimental constraints prevent us from modifying it.

\subsection{Violations of the equivalence principle and de-gravitating
  the vacuum}

Perhaps general relativity can be modified selectively, so that only
vacuum fluctuations do not couple to gravity?  Empirically we know
that virtual particles contribute to the inertial mass, for example
through the Lamb shift.  But perhaps the equivalence principle is
violated, and they do not contribute to the gravitational mass?

In fact, free fall experiments show that virtual particles do
gravitate in matter, satisfying the equivalence principle at least to
1 part in a million~\cite{Pol06}.  The only remaining possibility,
then, is to arrange that they gravitate in matter, but not in the
vacuum.  Ref.~\cite{Pol06} contains a careful discussion of the
difficulties of this approach, which I will not repeat here.

\subsection{Initial conditions}

Another tempting rug under which to sweep the cosmological constant
problem is the beginning of the universe.  Singularity theorems
suggest that classical spacetime had a beginning.  If so, there should
be a theory of initial conditions.  Perhaps it determines that the
universe must start out with zero vacuum energy?

In fact, this would be a disaster.  As discussed earlier, the energy
of the vacuum dropped sharply during various known phase transitions
in the early universe.  If the universe had started with
$\rho_\Lambda=0$, the vacuum energy would have decreased to
$-10^{-67}$ at the electroweak phase transition, leading to a big
crunch.  The universe would have ended when it was only $10^{-10}$
seconds old.  

This argument could also be used against the idea that a dynamical
mechanism attracted $\rho_\Lambda$ to 0 in the early universe.  But
attractor mechanisms are already ruled out by more general arguments,
which I will turn to next.

\subsection{Nongravitational dynamical attractor mechanisms}

Perhaps there is a local dynamical mechanism that allowed the standard
model to adjust coupling constants, masses, and/or effective
potentials, until their various contributions to the vacuum energy
cancelled out?  But nongravitational physics depends only on energy
differences, so the standard model cannot respond to the actual value
of the cosmological constant it sources.  This implies that
$\rho_\Lambda=0$ is not a special value from the particle physics point of
view.  In particular, it cannot be a dynamical attractor in a
nongravitational theory.

\subsection{Gravitational dynamical attractor mechanisms}

Gravity can see the cosmological constant, and among vacuum solutions
of Einstein's equation, Minkowski space ($\rho_\Lambda=0$) certainly
is special for having no curvature.  Perhaps, then, a dynamical
mechanism operated in the past, which attracted $\rho_\Lambda$ to zero
through gravitational interactions?  

The catch with this idea is that the universe is not vacuous.  Today's
cosmological constant was dynamically irrelevant in the early
universe.  This is one of the greatest difficulties in solving the
cosmological constant problem, and it is frequently overlooked.  A
mechanism that works only in an empty universe solves nobody's
problem.

It is worth going over this point in more detail.  Gravity couples to
the stress tensor, and vacuum energy is only one of many contributions
to the stress tensor.  Today, the energy density of the vacuum is
comparable to the average density of matter, and its pressure is
$10^4$ times greater than that of radiation.  But while matter and
radiation redshift under expansion, vacuum energy does not dilute.
Until the recent past, therefore, a vacuum energy density of
$10^{-120}$ would have constituted a negligible contribution to the
stress tensor.  Gravity was responding to much larger energies and
pressures.  The notion of a gravitational feedback mechanism adjusting
the cosmological constant to precision $10^{-120}$ in the early
universe can be roughly compared to an airplane following a prescribed
flight path to atomic precision, in a storm.

Consider, for example, the era of nucleosynthesis, at a temperature of
$1$ MeV.  The energy density of radiation was $\rho_{\rm
  BBN}=1.6\times 10^{-88}$ and its pressure was $p_{\rm
  BBN}=0.53\times 10^{-88}$.  Hence, no dynamical mechanism could have
adjusted the vacuum energy to precision better than $\delta
\rho_\Lambda \approx 10^{-88}$ prior to nucleosynthesis.  This exceeds
the conservative upper bound of Eq.~(\ref{eq-small}) by a factor of
$10^{33}$.

This problem can be restated in a geometric language.  Spacetime
geometry is the only physical entity that {\em can\/} be affected by
vacuum energy, but it need not be, because other forms of energy might
curve it more.  At nucleosynthesis, the curvature scale was $H_{\rm
  BBN}^{-1}\approx 3\times 10^{43}$.  {\em Any\/} cosmological constant
much smaller than $3H_{\rm BBN}^2/8\pi$ would have left this geometry
unaffected, so no imprint could have distinguished, say,
$\rho_\Lambda=10^{-90}$ from $\rho_\Lambda = 10^{-123}$ at that time,
and nothing could have selected for the latter.

One may speculate that a dynamical mechanism was operating
continuously, keeping the vacuum energy density comparable to the
density of matter or radiation at all times.  But this is impossible.
The cosmological dilution of a perfect fluid is completely determined
by the conservation of the stress tensor (and thus, in particular, by
general relativity).  Its density must redshift as $\rho=a^{-3(1+w)}$,
where $a$ is the scale factor and $p=w\rho$ is the pressure.  Hence,
vacuum energy, or anything behaving like it ($w\approx -1$) cannot
continuously dilute in sync with matter ($w=0$) or radiation
($w=1/3$).  For an analysis ruling out the continuous transfer of
vacuum energy into matter or radiation, see Ref.~\cite{Wei89}.

\subsection{Summary} I make no claim to have represented all
approaches one might try, nor have I identified all of their problems.
In order to focus on general constraints, I have granted various
optimistic assumptions and ignored technical gaps in several of the
ideas considered.

The obstructions listed here are powerful, but since some solution
must exist, they cannot be insurmountable.  It is best to think of
them as litmus tests.  A serious proposal should be able to state
precisely how it satisfies each constraint.  We have innumerable
``solutions'' that work in a world without matter and radiation, are
spoiled by symmetry breaking in the early universe, or appeal
exclusively to quantum gravity effects.  They fail to confront the
main difficulties of the problem.

\section{An old idea that might have worked but didn't}
\label{sec-nowdead}

The entire discussion so far could have been written in 1980.  The
cosmological constant problem---that $\rho_\Lambda$ is much smaller
than predicted---was well known then.  What was not known is whether
the cosmological constant is strictly zero or just very small, but
either way there was a crisis.  Undoubtedly, all of the above ideas
were already considered then, and were dismissed, since their flaws
are fatal quite independently of the precise value of
$\rho_\Lambda$.\footnote{Happily, this seems not to stand in the way
  of their continued, enthusiastic rediscovery.}

In this section I will discuss a strategy that would have been viable
in 1980 but has since become unattractive.  In order to give the idea
a fair hearing, let us ignore for a moment the 1998 discovery of a
nonzero cosmological constant.  We know only that
$|\rho_\Lambda|\lesssim 10^{-121}$, and we would like to understand
why.

\subsection{$\rho_\Lambda$ vanishes in the asymptotic future}

Except in the case of attractor behavior, the state of a dynamical
system cannot be determined from its equations of motion alone; one
needs some knowledge of boundary conditions.  But from
Sec.~\ref{sec-doa} we already know that neither initial conditions,
nor attractor behavior in the early universe, are likely to solve the
problem.

However, final conditions might work.  In the absence of a reliable
theory of either initial or final conditions, it is legitimate to
speculate that the boundary conditions on the universe are most
naturally formulated in the asymptotic future, and that they dictate
that $\rho_\Lambda=0$ in the late time limit.  (The final conditions
might also set $\rho_\Lambda$ to a small nonzero value, but this is
less attractive, since it introduces an arbitary scale.\footnote{A
  nonzero value, no matter how small, might even spoil the idea.
  Negative $\rho_\Lambda$ leads to a big crunch with high energy
  densities and perhaps also the restoration of some broken
  symmetries.  Setting $|\rho_\Lambda|$ to a small value at the big
  crunch thus faces one or more of the problems that frustrated
  attempts to fix it in the early universe.  Positive $\rho_\Lambda$
  at $t=\infty$ implies an eternal de~Sitter universe.  Under
  plausible assumptions, our observation of a universe with a long
  semiclassical history is ruled out in such a
  cosmology~\cite{DysKle02}; see, however, Ref.~\cite{Ban07}.})

What does this imply for the cosmological constant today?  All known
effective scalars are presently in their vacuum.  One can hypothesize
additional scalars that are not (such as quintessence), but this is
quite difficult to implement without significant fine-tuning, and in
any case constitutes a complication of the model.  In any natural
implementation of our idea, therefore, there will be no difference
between the vacuum energy today and the vacuum energy in the infinite
future, $\rho_\Lambda=0$.

\subsection{Predictions}

Thus, the final-condition approach generically predicts that there is
no vacuum energy in the present era.  This prediction has been
falsified.  Experiments began to show around 1998 that the vacuum
energy is positive and of order $10^{-123}$ (see Sec.~\ref{sec-duck}).

The final-condition approach makes a second prediction, which should
not be overlooked just because it is theoretical.  I made two very
strong assumptions:
\begin{enumerate}
\item The boundary conditions of the universe are set in the far
  future.
\item They require $\rho_\Lambda=0$.
\end{enumerate}
If the idea is right, these assumptions should eventually be
vindicated by progress in our understanding of fundamental theory.  

To date, at least, this has not happened.  Another way to say this is
to note that since the time when many theorists began worrying
seriously about the cosmological constant problem (the 1980s at the
latest), we have not succeeded in making this idea any more precise,
or in showing that it can actually be realized in a plausible context
such as string theory.  In summary, the experimental developments of
recent decades strongly disfavor this approach, while theoretical
developments have failed to corroborate it.

But in fact, the same can be said for other approaches that predicted
that the energy of the vacuum should vanish in the present era.  In
several decades of theoretical work, we have not identified any
concrete reason why it should; and since 1998 we know that it does
not.

\section{An old idea that could have been falsified but wasn't}
\label{sec-notyetdead}

I will now consider a different strategy, various versions of which
were suggested throughout the 1980s.  Its initial status was very
similar to the previous strategy: It made an experimental prediction
for $\rho_\Lambda$, and it optimistically anticipated that theoretical
progress would eventually justify a number of implicit assumptions.
Unlike the previous strategy, however, this one has since found
support on both counts.

\subsection{$\rho_\Lambda$ is an environmental variable}
\label{sec-var}

The strategy~\cite{Lin84c,Sak84,Ban85,Lin87,Wei87} is to
posit that the universe is much larger than its presently visible
portion, and that $\rho_\Lambda$ varies from place to place, though it
can be constant over very large distances.  According to
Weinberg~\cite{Wei87} (see also Refs.~\cite{DavUnw82,BarTip}),
structure such as galaxies will only form in locations where
\begin{equation}
  -10^{-123}\lesssim \rho_\Lambda\lesssim  3\times 10^{-121}~.
\label{eq-weinberg}
\end{equation}
Assuming that structure is a prerequisite for the existence of
observers, we should then not be surprised to find ourselves in such a
region.

Why is $\rho_\Lambda$ related to structure formation?  To form galaxies and
clusters, the tiny density perturbations visible in the cosmic
microwave background radiation had to grow under their own gravity,
until they became nonlinear and decoupled from the cosmological
expansion.  This growth is logarithmic during radiation domination,
and linear in the scale factor during matter domination.  

Vacuum energy does not get diluted, so it inevitably comes to dominate
the dynamics of the universe, at a time of order $t_\Lambda\sim
|\rho_\Lambda|^{-1/2}$.  If $\rho_\Lambda>0$, small perturbations will cease to
grow at this time.  The only structures that will remain are highly
overdense regions that have already become gravitationally bound and
decoupled from the cosmological expansion.  

This means that there would be no structure in the universe if the
cosmological constant had been large enough to dominate the energy
density before the first galaxies formed~\cite{Wei87}, tens or
hundreds of millions of years after the big bang.  Careful analysis
then leads to the upper bound in
Eq.~(\ref{eq-weinberg}).\footnote{Since this argument was first
  proposed, dwarf galaxies have been discovered at higher redshift.
  This raises the upper bound on $\rho_\Lambda$ obtained by Weinberg's
  argument, and it can make the observed value seem surprisingly small
  (though by a factor of $10^{-3}$, still much better than
  $10^{-123}$).  This discrepancy may grow if parameters other than
  $\Lambda$ can also to vary.  Its magnitude, however, depends on the
  manner in which the divergent numbers of observers living in
  different parts of the universe are regulated and compared.  The
  discrepancy is entirely absent in one natural proposal (see
  Sec.~\ref{sec-outlook}).} The lower bound comes about because the
universe would have already recollapsed into a big crunch if
$(-\rho_\Lambda)$ had been larger than the matter density
today~\cite{BarTip,Wei89}.

Note that Weinberg's point was not to improve the experimental upper
bound on $\rho_\Lambda$ but to explain its smallness.  In fact, the
limits on $\rho_\Lambda$ from galaxy formation are more lenient than
bounds derived from other data available at the time, such as
constraints on the expansion rate and flatness of the universe.  But
those data were not in any obvious way required for the existence of
observers, so they could not have been used in an anthropic
argument.

\subsection{Experimental Predictions}
\label{sec-wep}

If it is true that all observers live in regions satisfying
Eq.~(\ref{eq-weinberg}), then it is obvious why we do not observe a
value of $|\rho_\Lambda|$ greater than $10^{-120}$.  But what {\em
  do\/} we expect to see?  

We assumed that many different values of $\rho_\Lambda$ are possible,
but this does not mean they are all equally likely.  However,
Eq.~(\ref{eq-weinberg}) represents an exceedingly small interval
compared with the natural scale of $\rho_\Lambda$ (the latter being
unity, or in any case not less than $10^{-60}$).  It is plausible that
the likelihood of different values of $\rho_\Lambda$ should not vary
significantly over such a tiny interval.  (You may think that a
divergence at $\rho_\Lambda=0$ is possible, since 0 looks like a
special value, but I argued earlier that it is not.)  Technically,
this means that we should consider a ``flat'' prior probability for
values in Eq.~(\ref{eq-weinberg}):
\begin{equation}
dp/d\rho_\Lambda\approx~\mbox{const.}
\label{eq-flat}
\end{equation}

Then it would be surprising if we should find ourselves in a region
with, say, $\rho_\Lambda=10^{-150}$.  This would not be a typical value; it
would be much smaller than necessary for observers, an unlikely
accident.  It is far more likely that the local cosmological constant
has a {\em typical\/} value in the range compatible with structure
formation, which by Eqs.~(\ref{eq-weinberg}) and (\ref{eq-flat}) is of
order $10^{-121}$.

Therefore, the ``environmental'' approach predicts~\cite{Wei87} that
the vacuum energy should be nonzero and not much smaller than
$10^{-121}$.  In other words, its magnitude should be comparable to,
or somewhat larger than, the present matter density, $\rho_{\rm
  matter}\approx 4.6\times 10^{-124}$.  But this means that it should
be detectable by careful experiments.

Weinberg's prediction was confirmed in 1998, when it was discovered
that $\rho_\Lambda\approx 1.5\times 10^{-123}$.  I discuss this
development further in Sec.~\ref{sec-duck}.

\subsection{Theoretical Predictions}
\label{sec-wtp}

The environmental strategy makes several highly nontrivial
assumptions:
\begin{enumerate}
\item{$\rho_\Lambda$ is fundamentally not fixed but variable.}
\item{Its possible values are continuous, or are sufficiently closely
    spaced that Eq.~(\ref{eq-weinberg}) is satisfied by at least one
    of them.}
\item\begin{enumerate}
  \item{Either, boundary conditions ensure that $\rho_\Lambda$ will satisfy
      Eq.~(\ref{eq-weinberg}) in the present epoch}.
  \item{Or, starting from generic initial conditions, many other
      values are eventually realized in different spacetime regions by
      some dynamical mechanism.  In particular, at least one value of
      $\rho_\Lambda$ satisfying Eq.~(\ref{eq-weinberg}) can be dynamically
      attained somewhere.}
\end{enumerate}
\item{Regions satisfying Eq.~(\ref{eq-weinberg}) can grow larger than
    the entire observable universe, and can survive for longer than 13
    billion of years.}
\item{Such regions can contain matter and radiation.}
\end{enumerate}
Thus, the credibility of the environmental approach rested not only on
its (successful) experimental prediction.  It also depended on the
ability of future theoretical developments to provide a justification
for each of the assumptions involved.  

These theoretical predictions are not trivial.  It was far from clear
whether they could be satisfied and explained by a concrete model,
even at the level of effective field theory.  To see how hard it is,
it helps to consider one of the constructions that came closest (see
Sec.~\ref{sec-bt}).

String theory is particularly rigid in restricting the ingredients it
allows us to work with.  It might have seemed overwhelmingly unlikely,
therefore, that string theory should fit snugly into the complicated
mold of constraints described above---constraints that even less
fundamental, more flexible frameworks seemed unable to conform to.

Yet, we now have strong evidence that string theory succeeds at this
task.  The solution, the gist of which I will describe in Sections
\ref{sec-bp1} and \ref{sec-bp2}, depends crucially on some of
string theory's most characteristic, defining elements, such as the
existence of extra dimensions and of higher-dimensional objects
(branes).  This development is perhaps no less remarkable than the
experimental confirmation of Weinberg's prediction.

\section{Why dark energy is a cosmological constant}
\label{sec-duck}

We now know that the cosmological constant is not zero.  This was
discovered in 1998 by measuring the apparent luminosity of distant
supernovae~\cite{Rie98,Per98}.  Their dimness indicates that the
expansion of the universe has recently begun to accelerate, consistent
with a positive cosmological constant~\cite{Teg06}
\begin{equation}
  \rho_\Lambda= (1.48\pm 0.11) \times 10^{-123}~,
\label{eq-duck}
\end{equation}
and inconsistent with $\rho_\Lambda =0$.  Cross-checks have
corroborated this conclusion.  For example, the above value of
$\rho_\Lambda$ also explains the observed spatial flatness of the
universe, which cannot be accounted for by baryonic and dark matter
alone.  (See Ref.~\cite{Teg06} for a recent summary of constraints
from various experiments.)

What does this observation imply for the cosmological constant
problem?  It neither creates it nor solves it.  But it sharpens it,
and so discriminates powerfully between approaches to its
solution. The fact that $0\neq\rho_\Lambda\approx 3.2\,\rho_{\rm
  matter}$ disfavors theories that leads to vanishing $\rho_\Lambda$,
and it favors any theory that predicts $\rho_\Lambda$ to be comparable
to the present matter density.

\subsection{Calling it a duck}

I have failed to describe the recent discovery in tones of wonder and
stupefaction, as a ``mysterious dark energy'', a nonclustering fluid,
with equation of state $w=p_{\rm DE}/\rho_{\rm DE}$ close to $-1$,
which currently makes up 75\% of the energy density of the universe.
Why obfuscate?  If a poet sees something that walks like a duck and
swims like a duck and quacks like a duck, we will forgive him for
entertaining more fanciful possibilities.  It could be a unicorn in a
duck suit---who's to say!  But we know that more likely, it's a duck.

In science, it can be wrong to keep an open mind, and the expression
``dark energy'' is an example of misplaced political correctness.
Dark energy is the cosmological constant until proven otherwise, for
the same reason that the moon is not made of cheese until proven
otherwise: It is by far the most economical interpretation of the
data, even if it fails to sustain a fondly held preconception---in
this case, the prejudice that $\rho_\Lambda =0$.

Let me make this completely explicit.  A conservative, well-tested
framework, the standard model coupled to general relativity, has
encountered a problem: Why is $\rho_\Lambda $ much smaller than
several known contributions to it?  No proposal for its resolution can
claim a solid footing.\footnote{I will argue in the next section that
  this is no longer entirely true.  Still, none is on a footing
  comparable to the standard model or general relativity.}  If
experiment, rather than theoretical bias, is to be our guide, then we
must remain agnostic as to how the cosmological constant problem will
be solved.

Thus, $\rho_\Lambda $ remains an unexplained parameter that we must
fit to the data until help arrives.  Dark energy is experimentally
indistinguishable from vacuum energy, but definitely distinct from any
other known form of matter.  It is reasonable, then, to consider dark
energy to be vacuum energy, and to fit $\rho_\Lambda$ to its observed
density.

Thus, Eq.~(\ref{eq-duck}) is a straightforward, unbiased
interpretation of dark energy.  Without interposing theoretical
speculations, we have allowed experiment to determine precisely {\em
  which problem} physics (properly defined to include only
well-established and tested theories) actually faces: Why is
$\rho_\Lambda$ much smaller than many known contributions to it, and
why is it comparable to the energy density of matter today?

Only now may we turn to the realm of speculation, and ask {\em how the
  problem might be resolved}.  It behooves us to use experiment to
evaluate our hypotheses, not the other way around.  Among hypotheses
that made specific predictions for the value of $\rho_\Lambda$, we may
safely conclude that some (Sec.~\ref{sec-notyetdead}) are favored by
the discovery of nonzero $\rho_\Lambda$, and others
(Sec.~\ref{sec-nowdead}) are disfavored.

\subsection{Two problems for the price of one}

In order to motivate a different interpretation of ``dark energy'', we
would need to turn the scientific method on its head, and begin by
decreeing that the cosmological constant problem will one day be
resolved by a theory that sets $\rho_\Lambda =0$ precisely.  We would
have to treat this claim not as a hypothesis among others, to be
judged against empirical evidence, but as a dogma that constrains our
interpretation of any observation.

As theoretical bets go, this one is daring: No concrete, viable theory
predicting $\rho_\Lambda=0$ was known by 1998, and none has been found
since.  By contrast, we do have a concrete implementation of
Weinberg's approach, which predicts $\rho_\Lambda \sim \rho_{\rm
  matter}$.  But this will be the topic of Sec.~\ref{sec-theory}.

Here, my argument is not about which bet we should make.  We should
not be betting at all.  By looking at the world through the lens of
just one hypothesis, we prevent experiment from discriminating between
it and other hypotheses, depriving ourselves of the fruits of
considerable labor.

To make matters worse, the assumption that $\rho_\Lambda=0$ is not
just pure speculation.  It is pure speculation that turns one problem
into two:
\begin{enumerate}

\item Why is $\rho_\Lambda =0$?\footnote{This problem is sometimes
    overlooked, as if $\rho_\Lambda=0$ required no
    explanation~\cite{DETF}---a potentially expensive fallacy that
    greatly exaggerates the plausibility of time-dependent dark
    energy.  It creates false expectations and may distort our
    assessment of important future experiments.}  (This is the old
  cosmological constant problem of Sec.~\ref{sec-ccp}, made more
  specific not by experiment but by our decree that $\rho_\Lambda=0$.)

\item If, by our decree, dark energy is not vacuum energy, then it
  must be something else---but something that looks and acts, in every
  respect so far observed, just like vacuum energy, and unlike any
  other known substance.  What could it be?

\end{enumerate}
The two problems have very different origins.  The cosmological
constant problem, at least in its original form, is real, in that it
arises within well-tested physics.  The dark energy problem is an
artifact produced by insisting on the untested hypothesis that
$\rho_\Lambda=0$.

On the upside, much progress is being made on the second problem (if
we count fine-tuned scalar fields and {\em ad hoc\/} modifications of
gravity\footnote{Infrared modifications of gravity do not solve the
  cosmological constant problem (see Sec.~\ref{sec-doa}), but given
  enough small parameters, they can mimic dark energy.}).  Ironically,
the very observations that have finally provided a clue about the
cosmological constant problem---and which, for that reason alone,
would rank among the great triumphs of experimental science---seem to
have diverted our attention to the entirely fictitious problem of dark
energy.

\subsection{The real second problem}
\label{sec-coincidence}

By sharpening the cosmological constant problem, the discovery of
nonzero vacuum energy did create a new challenge, sometimes
called the coincidence problem or why-now problem.  Vacuum energy, or
anything behaving like it (which includes all options still allowed by
current data) does not redshift like matter.  In the past, vacuum
energy was negligible, and in the far future, matter will be very
dilute and vacuum energy will dominate completely.  The two can be
comparable only in a particular epoch.  It is intriguing that this is
the same epoch in which we are making the observation.

Note that this apparent coincidence involves us, the observers, in an
essential way.  If it has any explanation, it will by definition have
to be an anthropic explanation.

In fact, Weinberg's approach (Sec.~\ref{sec-notyetdead}) explained the
coincidence before it was discovered.  It predicts that vacuum energy
will be just small enough to allow galaxies to form, which implies
that in the epoch immediately following galaxy formation it will be
comparable to the matter density.  This explains the coincidence if
one assumes not only that observers require galaxies, but also that
typical observers form not too long after galaxies do.  (Both
assumptions seem to hold in our universe, but in the context of the
string landscape their general validity could be debated.  At the end
of Sec.~\ref{sec-outlook}, I will discuss a natural measure on the
multiverse that renders both assumptions unnecessary, explaining the
coincidence more directly and more generally.)

\section{A theory of the cosmological constant}
\label{sec-theory}

In this section I review a theory of the cosmological constant that
realizes the general idea of Sec.~\ref{sec-notyetdead}.  It satisfies
all five constraints identified in Sec.~\ref{sec-wtp}, and it arises
naturally in string theory.

I will begin in Sec.~\ref{sec-bt} by reviewing an older idea.  The
model does not succeed, but it presents a natural starting point for
the discussion of a more powerful mechanism in Sec.~\ref{sec-bp1}.  In
Sec.~\ref{sec-bp2}, I argue that this construction arises naturally in
string theory.

\subsection{The Brown-Teitelboim mechanism}
\label{sec-bt}

In this subsection, I will discuss an influential construction by
Brown and Teitelboim~\cite{BT1,BT2} that anticipated a number of
features of the landscape of string theory.  It will turn out to
satisfy a subset of the conditions (1)-(5) of Sec.~\ref{sec-wtp},
bringing the remaining challenges into sharp profile.

\paragraph{Discretely adjustable vacuum energy from a four-form
  field}

Extending an idea of Abbott~\cite{Abb85}, Brown and
Teitelboim~\cite{BT1,BT2} introduced a four-form field to make the
cosmological constant variable (condition 1).  An analogy with
electromagnetism helps clarify its role.

The Maxwell field, $F_{ab}$, is derived from a potential, $F_{ab}
= \partial_a A_b - \partial_b A_a$.  The potential is sourced by a
point particle through a term $\int e \mathbf A$ in the action, where
the integral is over the worldline of the particle, and e is the
charge.  Technically, $\mathbf F$ is a two-form (a totally
antisymmetric tensor of rank 2), and $\mathbf A$ is a one-form
coupling to a one-dimensional worldvolume (the worldline of the
electron).

The field content of string theory and supergravity is completely
determined by the structure of the theory.  It includes a four-form
field, $F_{abcd}$, which derives from a three-form potential:
\begin{equation}
F_{abcd} = \partial_{[a} A_{bcd]}~,
\end{equation}
where square brackets denote total antisymmetrization.  This potential
naturally couples to a two-dimensional object, a membrane, through a
term $\int q \mathbf A$, where the integral runs over the 2+1
dimensional membrane worldvolume, and $q$ is the membrane charge.

The properties of the four-form field in our 3+1 dimensional world
mirror the behavior of Maxwell theory in a 1+1 dimensional system.
Consider, for example, an electric field between two parallel
capacitor plates of equal and opposite charge.  If the plates are very
large, then the field strength in the interior is constant both in
space and time.  Its magnitude depends on how many electrons the
negative plate contains; thus it will be an integer multiple of the
electron charge: $E = ne$.

The energy density will be one half of the field strength squared:
\begin{equation}
\rho = \frac{F_{ab}F^{ab}}{2} = \frac{n^2e^2}{2}
\end{equation}
In order to treat this as a system with only one spatial dimension, I
have integrated over the directions transverse to the field lines, so
$\rho$ is energy per unit length.  The pressure is equal to $-\rho$.
The corresponding 1+1 dimensional stress tensor has the form of
Eq.~(\ref{eq-lrho}), so the electromagnetic stress tensor acts like
vacuum energy in 1+1 dimensions.

The same is true for the four-form in our 3+1 dimensional world.
First of all, the equation of motion in the absence of sources is
$\partial_a(\sqrt{-g} F^{abcd}) = 0$, with
solution~\cite{DufNie80,AurNic80}
\begin{equation}
F^{abcd} = c \epsilon^{abcd}~,
\end{equation}
where $\epsilon $ is the unit totally anti-symmetric tensor and $c$ is
an arbitrary constant.  In string theory, there are ``magnetic''
charges (technically, five-branes) dual to the ``electric charges''
(the membranes) sourcing the four-form field.  Then, by an analogue of
Dirac quantization of the electric charge, one can show that $c$ is
quantized in integer multiples of the membrane charge, $q$:
\begin{equation}
c = nq~.
\end{equation}
Note that the actual value of the four-form field is thus quantized,
not only the difference between possible values~\cite{BP}.

The four-form field strength squares to $F_{abcd}F^{abcd} = 24 c^2$,
and the stress tensor is proportional to the metric, with
\begin{equation}
\rho = \frac{1}{2 \times 4!} F_{abcd}F^{abcd} = \frac{n^2 q^2}{2}
\end{equation}
In summary, the four-form field is nondynamical, and it contributes
$n^2 q^2/2$ to the vacuum energy.  It is thus indistinguishable from a
contribution to the cosmological constant.

\paragraph{Dynamics}

Classically, the field configurations we studied have no dynamics, but
quantum mechanically, they are unstable to nonperturbative tunneling
effects.  This is readily apparent in the electromagnetic, 1+1
dimensional analogy.  The electric field between the plates will be
slowly discharged by Schwinger pair creation of field sources.  This
is a process by which a electron and a positron tunnel out of the
vacuum.  Since field lines from the plates can now end on these
particles, the electric field between the two particles will be lower
by one unit [$ne\rightarrow (n-1)e$].  The particles will appear
precisely at such a separation that the corresponding decrease in
field energy compensates for their combined rest mass.  They are then
subjected to constant acceleration by the electric field until they
hit the plates.  If the plates are far away, they will move
practically at the speed of light by that time.

For weak fields, this tunneling process is exponentially suppressed,
with a rate of order $\exp(-\pi m^2/ne^2)$, where the exponent arises
as the action of a Euclidean-time solution describing the appearance
of the particles.  Thus, a long time passes between creation events.
However, over large enough time scales, the electric field will
decrease by discrete steps of size $e$.  Correspondingly, the 1+1
dimensional ``vacuum energy'', i.e., the energy per unit length in the
electric field, will gradually decrease by discrete amounts $[n^2
e^2-(n-1)^2e^2]/2 = (n-\frac{1}{2})e^2$.  Note that this step size
depends both on the electric charge, $e$, and on the remaining flux,
$n$.  The cascade of decays will only terminate once the electric
field has been depleted to the point where not enough energy is left
for the creation of another electron-positron pair.

Precisely analogous nonperturbative effects occur for the four-form
field in 3+1 dimensions.  By an analogue of the Schwinger process,
spherical membranes can spontaneously appear.  (This is the correct
analogue: the two particles above form a zero-sphere, i.e., two
points; the membrane forms a two-sphere.)  Inside this source, the
four-form field strength will be lower by one unit of the membrane
charge [$nq\rightarrow (n-1)q$].  The process conserves energy: the
initial membrane size is such that the membrane mass is balanced
against the decreased energy of the four-form field inside the
membrane.  The membrane quickly grows to convert more space to the
lower energy density, accelerating outward and expanding
asymptotically at the speed of light.

Membrane creation is a well-understood process described by a
Euclidean instanton, and like Schwinger pair creation, is generically
exponentially slow.  Ultimately, however, it will lead to the
step-by-step decay of the four-form field.  Inside a new membrane, the
vacuum energy will be lower by\footnote{Here and below, I am using
  reduced Planck units, obtained by replacing $G$ with $8\pi G$ in the
  definition of all Planck units.  This avoids annoying factors of
  $8\pi$ (at the expense of consistency with earlier sections).  In
  the new units, $\Lambda=\rho_\Lambda$; Weinberg's upper bound in
  Eq.~(\ref{eq-weinberg}) is $2\times 10^{-118}$; and the observed
  value corresponding to Eq.~(\ref{eq-duck}) is $0.9\times
  10^{-120}$.}
\begin{equation}
\delta\Lambda=(n-\frac{1}{2})q^2~.
\label{eq-deltalambda}
\end{equation}

This suggests a mechanism for cancelling off the cosmological
constant.  Let us collect all contributions (see Sec.~\ref{sec-ccp}),
except for the four-form field, in a ``bare'' cosmological constant
$\lambda$.  Generically, $|\lambda|$ should be of order unity (at
least in the absence of supersymmetry), and we will assume without
excessive loss of generality that it is negative.  With $n$ units of
four-form flux turned on, the full cosmological constant will be given
by
\begin{equation}
\Lambda = \lambda+ \frac{1}{2} n^2 q^2
\end{equation}
If $n$ starts out large, the cosmological constant will decay by
repeated membrane creation, until it is close to zero.  

\paragraph{Limitations}

However, we must verify whether $\Lambda$ gets close {\em enough\/} to
zero, i.e., whether condition (2) can be satisfied.  The smallest
value of $|\Lambda|$ is attained for the flux $n_{\rm best}$, given by
the nearest integer to $\sqrt{2|\lambda|}/q$. The step size near
$\Lambda=0$ is thus given by $(n_{\rm best}-\frac{1}{2})q^2$.  For the
Brown-Teitelboim mechanism to produce a value in the Weinberg window,
Eq.~(\ref{eq-weinberg}), this step size would need to be of order
$10^{-118}$ or smaller.  This requires an extremely small membrane
charge,
\begin{equation}
q\lesssim 10^{-118} |\lambda|^{-1/2}~.
\label{eq-smallcharge}
\end{equation}
A natural bare cosmological constant $\lambda$ will be no smaller than
of order $10^{-64}$, so $q\lesssim 10^{-86}$.

Such a small membrane charge $q$ is unnatural.  In particular, it is
not known how to realize a sufficiently small charge in string theory.
Thus, it is not clear how condition (2) is to be satisfied.  This is
the {\em step size problem}.

This still looks like progress: Naively, it would seem that we have
succeeded in reducing the cosmological constant problem to a hierarchy
problem.  All we need is to introduce a small coupling and stabilize
it against corrections.  At least in principle, this is something we
know how to do, and perhaps the details can be worked out later.  In
fact, however, this example illustrates just how much worse the
cosmological constant problem is.  To see this, let us assume the
small-charge problem solved and take Eq.~(\ref{eq-smallcharge}) to be
satisfied.

In this hypothetical model, the Brown-Teitelboim adjustment mechanism
would indeed produce a small cosmological constant, but only in
regions containing no matter and radiation.  This is not surprising.
Rather general arguments in Sec.~\ref{sec-doa} showed that a
primordial dynamical adjustment mechanism for $\Lambda$ can only be as
accurate as the energy density in the early universe, which was much
larger than $10^{-118}$.  Now we encounter the other side of the same
coin: the {\em empty universe problem}.  We have found a mechanism
which cancels $\Lambda$ to high precision, but at the price of
removing all matter as well.\footnote{Steinhardt and
  Turok~\cite{SteTur06} consider an extension of another small-step
  model~\cite{Abb85}, aiming to overcome its empty universe problem.
  Their mechanism requires the universe to pass exponentially many
  times through an apparent big crunch/big bang transition, and to do
  so with negligible integrated probability of disturbing the
  small-step field governing the cosmological constant.  (These are
  strong assumptions whose validity remains to be demonstrated.)  With
  the additional assumption that the spacing of values of
  $\rho_\Lambda$ is of order $10^{-123}$ (the unsolved step size
  problem), this model would allow for the production of hot regions
  with small cosmological constant.  All other vacua, with larger
  cosmological constant, will also be produced in different spacetime
  regions; and every region, in most of its four-volume, is empty.
  Thus, anthropic localization in spacetime is necessary in any case
  to explain the observed $\rho_\Lambda$.  However, small-step
  single-field models like Refs.~\cite{Abb85,BT1,SteTur06} differ from
  a large-step, multi-field model like the string landscape in that
  nearly every worldline will experience {\em all\/} positive values
  of the cosmological constant, including the smallest one.
  Steinhardt and Turok argue that this is a desirable feature: ``All
  other things being equal, a theory that predicts that life can exist
  almost everywhere is overwhelmingly preferred by Bayesian analysis
  (or common sense) over a theory that predicts it can exist almost
  nowhere.''  At present, other things are far from equal, but I would
  not accept this particular criterion even as a tie-breaker.  It
  would suggest that a theory that predicts a universe densely packed
  with suns and earths is preferable to one that predicts large voids,
  where life cannot exist.  Put differently, we have already observed
  that most patches of space do not harbor life, so it seems
  questionable to demand that a good theory predict the opposite.  I
  find it more reasonable to judge a theory by whether it predicts
  correctly what observers observe, from economical and compelling
  assumptions.  Both in the observable universe, and in the
  multiverse, the dominance of empty regions is a dynamical
  consequence of a simple theory.  The fact that we do not live in
  such a region is not considered a problem, since it is an obvious
  consequence of the absence of matter.}

In detail the problem shows up as follows.\footnote{The empty universe
  haunts many other attempts to solve the cosmological constant
  problem, such as Ref.~\cite{Haw84b} and (I would argue in spite of
  Ref.~\cite{KleSus88b}) even Coleman's famous wormhole
  approach~\cite{Col88a}, which suffers in any case from technical
  problems.  In these cases it takes on a different guise: the
  amplitude of the wavefunction of the universe diverges for empty
  universes with vanishing cosmological constant.  Anthropic arguments
  will not help, because the probability for a universe containing
  {\em only\/} observers vastly exceeds the probability for the
  universe we see.}  Because membrane nucleation is a slow tunneling
process, small values of $\Lambda$ are approached very gradually from
above.  While the universe waits for the next membrane to nucleate, it
is dominated by positive vacuum energy.  Hence, it expands at an
exponential rate, rapidly diluting the energy density of any matter or
radiation that might have been around.  By the time a membrane appears
and the flux is reduced by one unit, the universe is completely empty.
This is true at every step, and so it will be true in particular when
the vacuum energy enters the Weinberg window.

Some energy is liberated when a membrane appears, because the vacuum
energy drops by $\delta\Lambda$.  Most of this energy goes into
accelerating the membrane as it expands outward.  (This is the famous
graceful-exit problem of old inflation.)  But let us be overly
optimistic and assume that instead all the energy goes into the
production of new matter and radiation.  Unfortunately, we were forced
earlier to assume that the step size is very small: each membrane
nucleation decreases the vacuum energy by $\delta\Lambda\sim
10^{-118}$ or less, or else our downward cascade would miss the
Weinberg window.  With $\rho\sim T^4$, this freed-up energy could at
best reheat the universe to a temperature $T\sim 10^{-30}$, or about
$10^{-2}$eV.  This falls well short of the $10$ MeV necessary to make
contact with standard cosmology, a theory we trust at least back to
nucleosynthesis.

In summary, the Brown-Teitelboim mechanism, with one four-form field,
fails to satisfy conditions (2) and (5) of Sec.~\ref{sec-wtp}.  The
exceedingly small step size required for a sufficiently dense spectrum
(2) cannot be attained in a natural model, and in any case the
associated slow descent would ensure that regions with small
cosmological constant are devoid of matter and radiation (5).

\subsection{The discretuum}
\label{sec-bp1}
  
The above problems can be overcome by considering a theory with more
than one species of four-form field~\cite{BP}.  In
Sec.~\ref{sec-bp2}, I will explain why this situation arises
naturally in string theory.  First let us see how multiple four-form
fields can produce a dense ``discretuum'' without requiring small
charges, and how this solves the empty universe problem.

\paragraph{Multiple four-form fields}

Consider a theory with $J$ four-form fields.  Correspondingly there
will be $J$ membrane species, with charges $q_1,\ldots,q_J$.  Above, I
analyzed the case of a single four-form field; essentially the
conclusions still apply to each field separately.  In particular, each
field strength separately will be constant in 3+1 dimensions,
\begin{equation}
F^{abcd}_{(i)} = n_i q_i \epsilon^{abcd}~,
\end{equation}
and it will contribute a discrete amount of vacuum energy to the
stress tensor.

Let us again collect all contributions to vacuum energy, except for
those from the $J$ four-form fields, in a bare cosmological constant
$\lambda$, which I assume to be negative but otherwise generic (i.e.,
of order unity).  Then the total cosmological constant will be given
by
\begin{equation}
\Lambda = \lambda + \frac{1}{2}\sum_{i=1}^J n_i^2 q_i^2~.
\label{eq-lmult}
\end{equation}
This will include a value in the Weinberg window,
Eq.~(\ref{eq-weinberg}), if there exists a set of integers $n_i$ such
that
\begin{equation}
2|\lambda|<\sum n_i^2 q_i^2<2(|\lambda|+ \Delta\Lambda)~,
\end{equation}
where $\Delta\Lambda\approx 10^{-118}$.  

A nice way to visualize this problem is to consider a $J$-dimensional
grid, with axes corresponding to the field strengths $n_i q_i$, as
shown in Fig.~\ref{fig-grid}.  
\EPSFIGURE{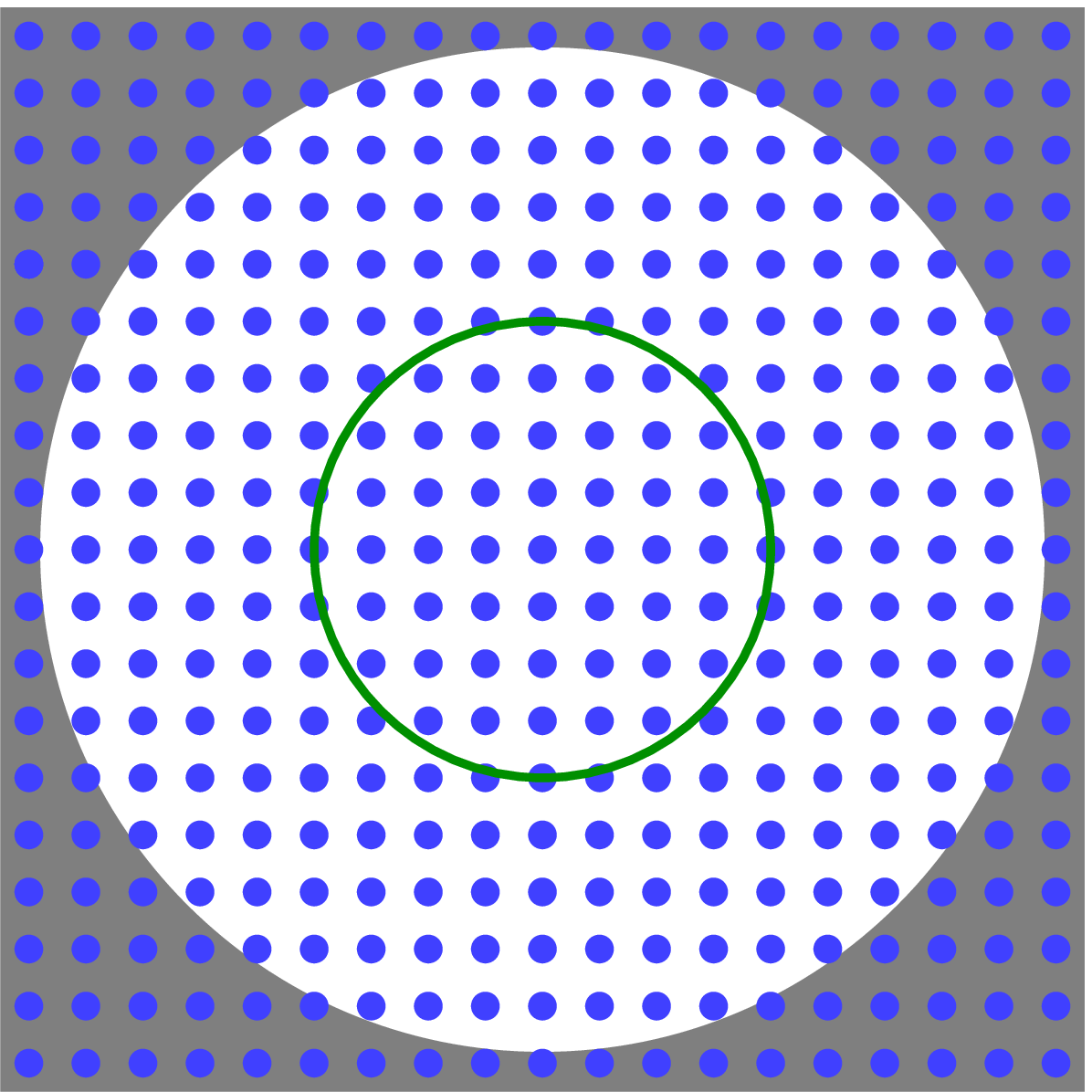,width=.6\textwidth}{\label{fig-grid} Possible
  configurations of the four-form fluxes correspond to discrete points
  in a $J$-dimensional grid, of which a two-dimensional section is
  shown.  The grid spacing in the $i$-th direction is the charge $q_i$
  of the corresponding membrane species. The Weinberg window
  corresponds to the thin (green) shell.  Inside the shell,
  $\Lambda<0$; outside $\Lambda>0$.  The physically relevant regime
  ($-1\lesssim \Lambda\lesssim 1$) is shown on white background.}
Every possible configuration of the four-form fields corresponds to a
list of integers $n_i$, and thus to a discrete grid point.  The
Weinberg window can be represented as a thin shell of radius
$\sqrt{2|\lambda|}$ and width $\Delta\Lambda/\sqrt{2|\lambda|}$.  The
shell has volume
\begin{equation}
V_{\rm shell} = \Omega_{J-1} (\sqrt{2|\lambda|})^{J-1}
\frac{\Delta\Lambda}{\sqrt{2|\lambda|}} = \Omega_{J-1}
|2\lambda|^{\frac{J}{2}-1}\Delta\Lambda~,
\end{equation}
where $\Omega_{J-1} = 2\pi^{J/2}/\Gamma(J/2)$ is the area of a unit
$J-1$ dimensional sphere.  The volume of a grid cell is
\begin{equation}
V_{\rm cell}=\prod_{i=1}^J q_i~.
\end{equation}
There will be at least one value of $\Lambda$ in the Weinberg window,
if $V_{\rm cell}<V_{\rm shell}$, i.e., if
\begin{equation}
\frac{\prod_{i=1}^J q_i}{\Omega_{J-1}
  |2\lambda|^{\frac{J}{2}-1}} < |\Delta\Lambda|~.
\label{eq-bp}
\end{equation}

The most important consequence of this formula is that charges no
longer need to be very small.  I will shortly argue that in string
theory one naturally expects $J$ to be in the hundreds.  With $J=100$,
for example, Eq.~(\ref{eq-bp}) [and thus, condition (2) of
Sec.~\ref{sec-wtp}] can be satisfied with charges $q_i$ of order
$10^{-1.6}$, or $\sqrt{q_i}\approx 1/6$ (the latter has mass dimension
1 and so seems an appropriate variable for the judging naturalness of
this scenario).  Interestingly, the large expected value of the bare
cosmological constant is actually welcome: it becomes more difficult
to satisfy Eq.~(\ref{eq-bp}) if $|\lambda|\ll 1$.

As it turns out, the largeness of the charges will also allow
condition (5) to be satisfied in a model with multiple four-form
fields: regions with small cosmological constant can contain matter
and radiation.

\paragraph{Dynamics}

Classically, every flux configuration ($n_1,\ldots,n_J$) is stable.
But quantum-mechanically, fluxes can change if a membrane is
spontaneously created.  As discussed in Sec.~\ref{sec-bt}, this
Schwinger-like process is generically exponentially suppressed.  Thus,
multiple four-forms naturally give a dense discretuum of metastable
vacua which can have extremely long life-times.

Starting from generic initial conditions, the universe will grow
arbitrarily large.  Over time, it will come to contain enormous
regions (``bubbles'' or ``pockets'') corresponding to each metastable
vacuum (Fig.~\ref{fig-global}).  In particular, our vacuum will be
realized somewhere in this ``multiverse''.  Moreover, it can be
efficiently reheated, so the empty-universe problem of
Sec.~\ref{sec-bt} will not arise.  Let us see how this works in more
detail.
\EPSFIGURE{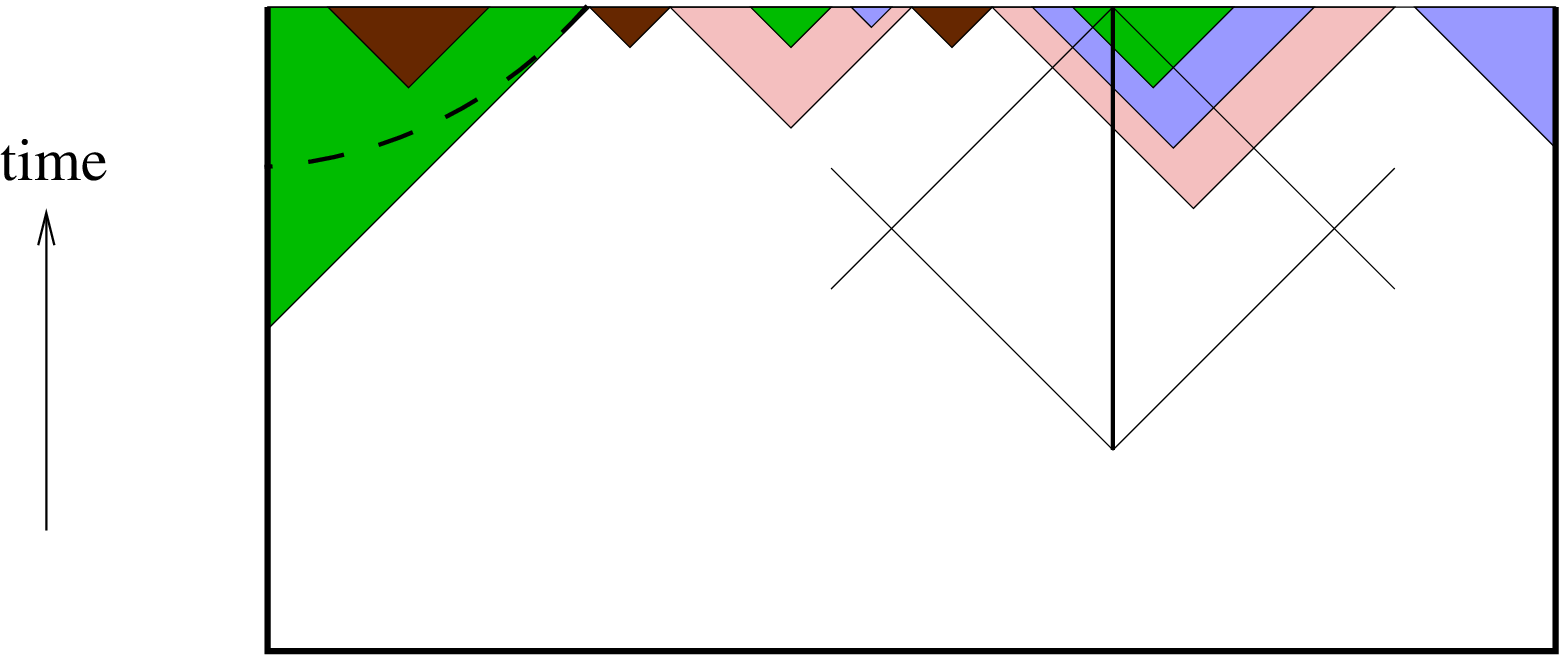,width=.8\textwidth}{\label{fig-global} Bird's
  eye view of the universe.  The triangles are pocket universes
  corresponding to different vacua in the landscape.  Each pocket is
  an infinite, spatially open universe; the dashed line shows an
  example of an instant of time as picked out by constant density
  hypersurfaces in the pocket.---This is a conformal diagram.  The
  actual amount of physical time and volume near the top boundary is
  infinite, and the top boundary is a fractal containing an infinite
  number of pocket universes.  The black diamond is an example of a
  spacetime region that is causally accessible to a single worldline
  (see Sec.~\ref{sec-outlook}).}

By Eq.~(\ref{eq-lmult}), all but a finite number of metastable vacua
will have $\Lambda>0$.  Let us assume that the universe begins in one
of these vacua.  Of course, this means that typically the cosmological
constant will be large initially.  Since $\Lambda>0$, the universe
will be well described by de~Sitter space.  It can be thought of as a
homogeneous, isotropic universe expanding exponentially on a
characteristic time scale $t_\Lambda\sim \Lambda^{-1/2}$.

Every once in a long while (this time scale being set by the action of
a membrane instanton, and thus typically much larger than
$t_\Lambda$), a membrane will spontaneously appear and the
cosmological constant will jump by $(n_i-\frac{1}{2}) q_i^2$.  But
this does not affect the whole universe.  $\Lambda$ will have changed
only inside the membrane bubble.  This region grows arbitrarily large
as the membrane expands at the speed of light.

But crucially, this does {\em not}\/ imply that the whole universe is
converted into the new vacuum~\cite{ColDel80}.  This technical result
can be understood intuitively.  The ambient, old vacuum is still, in a
sense, expanding exponentially fast.  The new bubble eats up the old
vacuum as fast as possible, at nearly the speed of light.  But this is
not fast enough to compete with the background expansion.

More and more membranes, of up to $J$ different types, will nucleate
in different places in the rapidly expanding old vacuum.  Yet, there
will always be some of the old vacuum left.  One can show that the
bubbles do not ``percolate'', i.e., they will never eat up all of
space~\cite{GutWei83}.  Thus different fluxes can change, and
different directions in the $J$-dimensional flux space are explored.

Inside the new bubbles, the game continues.  As long as $\Lambda$ is
still positive, there is room for everyone, because the background
expands exponentially fast.  In this way, all the points in the flux
grid $(n_1,\ldots,n_J)$, are realized as actual regions in physical
space.  The cascade comes to an end wherever a bubble is formed with
$\Lambda<0$, but this affects only the interior of that particular
bubble (it will undergo a big crunch).  Globally, the cascade
continues endlessly (Fig.~\ref{fig-global}).

Perhaps surprisingly, each bubble interior is an open FRW universe in
its own right, and thus infinite in spatial extent.\footnote{In an
  open universe, spatial hypersurfaces of constant energy density are
  three-dimensional hyperboloids.  This shape is dictated by the
  symmetries of the instanton describing the membrane nucleation.  It
  is closely related to the hyperbolic shape of the spacetime paths of
  accelerating particles, like the electron-positron pair studied
  above.} Yet, each bubble is embedded in a bigger universe (sometimes
called ``multiverse'' or ``megaverse''), which is extremely
inhomogeneous on the largest scales.

An important difference to the model with only one four-form is that
the vacua will not be populated in the order of their vacuum energy.
Because charges are large, two neighboring points in flux space will
differ enormously in cosmological constant.  That is, they differ by
one unit of flux, and the charges $q_i$ are not much smaller than one,
so by Eq.~(\ref{eq-lmult}) this translates into an enormous difference
in cosmological constant, of order $\delta\Lambda\sim q^2$ or more.
Conversely, vacua with very similar values of the cosmological
constant will be well separated in the flux grid, and will not
directly decay into one another.

This feature is crucial for solving the empty universe problem.  When
our vacuum was produced in the interior of a new membrane, the
cosmological constant may have decreased by as much as $1/100$ of the
Planck density.  Hence, the temperature before the jump was enormous
(in this example, the Gibbons-Hawking temperature~\cite{GibHaw77a} of
the corresponding de~Sitter universe would have been of order $1/10$
of the Planck temperature), and only extremely massive fields will
have relaxed to their minima.  Most fields will be thermally
distributed and can only begin to approach equilibrium after the jump
decreases the vacuum energy to near zero.

Thus, the final jump takes on a role analogous to the big bang in
standard cosmology.  The ``universe'' (really, just our particular
bubble) starts out hot and dense.  If the effective theory in the
bubble contains scalar fields with suitable potentials, there will be
a period of slow-roll inflation as their vacuum energy slowly relaxes.
(This was apparently the case in our vacuum.)  At the end of this
slow-roll inflation process, the universe reheats.

\subsection{String theory}
\label{sec-bp2}

It seems {\em ad hoc\/} to posit the existence of hundreds of species
of membranes, though perhaps a small price to pay for solving the
cosmological constant problem.  In fact, however, they arise
inevitably when string theory is applied to our four-dimensional
world~\cite{BP}.

\paragraph{Membrane species and extra dimensions}

The origin of the large number of four-form fields lies in the
topological complexity of small extra dimensions.  String theory is
formulated in 9+1 or 10+1 spacetime dimensions.  For definiteness, let
us work with the latter formulation (also known as M-theory).  If it
describes our world, then 7 of the spatial dimensions must be
compactified on a scale that would have eluded our most careful
experiments.  Thus one can write the spacetime manifold as a direct
product:
\begin{equation}
M = M_{3+1}\times X_7~.
\end{equation}
Typically, the compact seven-dimensional manifold $X_7$ will have
considerable topological complexity, in the sense of having large
numbers of noncontractable cycles of various dimensions.

To see what this will mean for the 3+1 dimensional description,
consider a string wrapped around a one-cycle (a ``handle'') in the
extra dimensions.  To a macroscopic observer this will appear as a
point particle, since the handle cannot be resolved.  Now, recall that
M-theory contains five-branes, the magnetic charges dual to membranes.
Like strings on a handle, five-branes can wrap higher-dimensional
cycles within the compact extra dimensions.  A five-brane wrapping a
three-cycle (a kind of noncontractible three-sphere embedded in the
compact manifold) will appear as a two-brane, i.e., a membrane, to the
macroscopic observer.

Six-dimensional manifolds, such as Calabi-Yau geometries, generically
have hundreds of different three-cycles, and adding another dimension
will only increase this number.  The five-brane---one of a small
number of fundamental objects of the theory---can wrap any of these
cycles, giving rise to hundreds of apparently different membrane
species in 3+1 dimensions, and thus, to $J\sim O(100)$ four-form
fields, as required.

The charge $q_i$ is determined by the five-brane charge (which is set
by the theory to be of order unity), the volume of $X_7$, and the
volume of the $i$-th three-cycle.  The latter factors can lead to
charges that are slightly smaller than 1, which is all that is
required.  Note also that the volumes of the three-cycles will
generically differ from each other, so one would expect the $q_i$ to
be mutually incommensurate.  This is important to avoid degeneracies
in Eq.~(\ref{eq-lmult}).

\paragraph{Vacuum stabilization}

The model I have presented is an oversimplification.  When it was
first proposed, it was not yet understood how to stabilize the compact
manifold against deformations (technically, how to give a mass to all
moduli fields including the dilaton).  This is clearly necessary in
any case if string theory is to describe our world, since we do not
observe massless scalars.  But one would expect that in a realistic
compactification, the fluxes wrapped on cycles should deform the
compact manifold, much like a rubber band wrapping a doughnut-shaped
balloon.  Yet, I have pretended that $X_7$ stays exactly the same
independently of the fluxes $n_i$.

Therefore, Eq.~(\ref{eq-lmult}) will not be correct in a realistic
model.  The charges $q_i$, and indeed the bare cosmological constant
$|\lambda|$, will themselves depend on the integers $n_i$.  Thus the
cosmological constant may vary quite unpredictably.  But the crucial
point remains unchanged: the number of metastable vacua, $N$, can be
extremely large, and the discretuum should have a typical spacing
$\Delta\Lambda\approx 1/N$.  For example, if there are $500$
three-cycles and each can support up to 9 units of flux, there will be
of order $N=10^{500}$ metastable configurations.  If their vacuum
energy is effectively a random variable with at most the Planck value
($|\Lambda|\lesssim 1$), then there will be $10^{380}$ vacua in the
Weinberg window, Eq.~(\ref{eq-weinberg}).

In the meantime, there has been significant progress with stabilizing
the compact geometry (e.g., Refs.~\cite{DasRaj99,GidKac01}; see
Refs.~\cite{Sil04,Gra05,DouKac06} for reviews.).  In particular,
Kachru, Kallosh, Linde, and Trivedi~\cite{KKLT} have shown that
metastable de~Sitter vacua can be realized in string theory while
fixing all moduli. (Constructions in noncritical string theory were
proposed earlier~\cite{Sil01,MalSil02}.)  These results confirmed the
above argument that the number of flux vacua can be large enough to
solve the cosmological constant problem.  More sophisticated counting
methods~\cite{DenDou04b} bear out the quantitative estimates obtained
from the simple model I have presented.

\section{Outlook: The landscape of string theory}
\label{sec-outlook}

The developments described in the previous section have changed the
status of the cosmological constant problem: we have, at last, a
concrete candidate for a solution.  Perhaps it is not the right
solution, but its existence makes it less acceptable to ignore the
problem, to fine-tune it away, or to indulge approaches that
demonstrably conflict with experiment.

They have also changed the status of string theory: the theory has
made contact with experiment---not merely in the sense of including a
smaller theory such as the standard model (which, arguably, can be
constructed from a suitable compactification), but in more
exhilarating ways: by being the first theory to explain a mysterious
observation that has long haunted us, and by doing so through means
completely its own.  Branes, fluxes, and extra dimensions are
inevitable in string theory.  They have turned out to be just what was
needed to get the job done.

And they have changed our thinking about how string theory will make
other predictions.  There are $10^{500}$ or more metastable vacua,
which can be thought of as local minima in a huge, multidimensional
potential landscape.  They differ not only in the value of their
vacuum energy, but in their entire low-energy effective field theory,
which is determined by local properties near the foot of a valley and
thus only very indirectly by the fundamental building blocks of the
landscape.  Different vacua will have different matter content,
coupling constants, and forces.  We will not predict the standard
model uniquely.  We will have to predict many of the features of our
universe statistically, from their relative abundance in the
landscape.

But it would be wrong to say that there are now $10^{500}$ ``string
theories'', suggesting a loss of fundamental simplicity and
uniqueness.  This is like saying that there are myriads of standard
models.  The standard model, like string theory, contains countless
metastable solutions, such as atoms, molecules, and condensed matter
at zero temperature.  They are all constructed from just five
different particle species (electrons, photons, and quarks).
Strictly, the number of possibilities is infinite; even with an energy
or volume cutoff, it quickly exceeds $10^{500}$.  This is not usually
considered a problem for the standard model.

Rather, a multitude of solutions is an essential feature of any theory
capable of explaining the multitude of complex phenomena that make up
the messy, real world.  It does not mean that anything goes.  There
are only a finite number of elements, and a random combination of
atoms is unlikely to form a stable molecule.  Even quantities such as
material properties ultimately derive from standard model parameters
and cannot be arbitrarily dialed.

The complexity of an object need not be an obstacle to its effective
description.  Suppose we set out to derive the properties of metals
from the standard model.  Looking at the size of the system, the
problem looks daunting, but we know very well that it yields to the
laws of large numbers.  Moreover, the predictive power of statistical
or effective theories is completely deterministic in practice.

The large number of vacua in string theory arises in a very similar
way, by combining a small set of fundamental ingredients in different
ways.  We cannot see the fluxes wrapped on handles in the extra
dimensions, but not long ago the same could be said about atoms, not
to speak of quarks.  We do not expect every aspect of a theory to be
testable, we just need to convince ourselves that it gives us more
than we put in.

So how do we get predictions from string theory?  Getting the
cosmological constant right is nice, but to confirm the landscape, we
need more.  We cannot create other vacua in the
laboratory~\cite{FarGut87}.  For now, all we can do is measure the
properties of our own, and do what we always do when we compare
experiment to theory: see if our observations are likely (i.e.,
typical~\cite{Pag07}), or unlikely, according to the theory.

This means making statistical predictions.  Since we cannot repeat
experiments in cosmology, the most interesting predictions will be
those that can be made with probability extremely close to 1, similar
to those in thermodynamics.  The problem can be divided into three
tasks, each of which are intensively studied at present.

\paragraph{Landscape statistics}
What is the relative abundance of stable or long-lived meta\-stable
vacua with specified low-energy properties?  This is the most obvious
question to ask in pursuit of statistically dominant features (see,
e.g., Refs.~\cite{Dou03,DenDou04b,GmeBlu05,DouTay06}, or
Ref.~\cite{Kum06} for a review).  Our understanding of metastable
vacua is still rather qualitative, so many investigations focus on
supersymmetric vacua instead, which are under better control.  It will
be important to develop more powerful techniques for dealing with
broken supersymmetry; meanwhile, it would help to understand the
extent to which current samples are representative of more realistic
vacua.

In particular, it can be useful to proceed by elimination.  One would
expect that most Lagrangians that make sense to a low-energy effective
field theorist cannot arise from the limited set of ingredients
supplied by string theory, no matter how elaborately they are
combined.  This vast ``swampland'' is not encompassed by the string
theory landscape.  The challenge is to identify specific predictions
that arise from such limitations~\cite{Vaf05,ArkMot06,OogVaf06}.

\paragraph{Dynamical selection effects}
Another major challenge is posed by eternal inflation, the
cosmological dynamics that produces different vacua in large, widely
separated regions of the universe. One would expect that this
mechanism favors some vacua over others, and thus enters into
statistical predictions.

Divergences in the global structure of the universe make this effect
very difficult to quantify.  As seen in Fig.~\ref{fig-global}, each
vacuum is realized infinitely many times as a bubble embedded in the
global spacetime.  Moreover, every bubble is an open universe and thus
of infinite spatial extent.

It would seem natural to regulate these infinities by considering the
universe at finite time before taking a limit.  However, there is an
ambiguity as to whether one should compare the volumes, or simply the
number of each type of bubble on this time slice (or some intermediate
quantity).  Either way, results depend strongly on the choice of time
variable~\cite{LinLin94,GarLin94}, which is rendered ambiguous by the
inhomogeneity of the global spacetime.
This is known as the measure problem of eternal inflation.  

A small number of relatively simple proposals arise by generating a
time variable from a geodesic congruence~\cite{%
  LinLin94,GarLin94,GarVil01,GarSch05,EasLim05,Lin06,Van06}, though
the fundamental significance of this procedure is not clear.  A more
radical proposal~\cite{Bou06}, motivated by black hole
complementarity, restricts attention to a spacetime region causally
accessible to a single worldline (a ``causal diamond''---see
Fig.~\ref{fig-global}).  In this case, the global distribution of
vacua (which is unobservable in any case~\cite{BouFre06a}) need not be
computed and regulated.  Instead, one computes the relative
probability that the worldline will enter a given vacuum.  This is
unambiguous and finite.

It is not yet clear how to derive the correct measure from first
principles (see Ref.~\cite{FreSus04} for an interesting approach).
But considerable progress has been made by the more pedestrian method
of elimination.  The number of simple candidate measures is not large,
and many make wrong predictions, which go by colorful names such as
Q-catastrophe, youngness paradox, Boltzmann brain paradox, and
staggering problem~\cite{%
  FelHal05,GarVil05,Pag06,BouFre06b,SchVil06,Lin06,Lin07,%
  BouYan07,OluSch07,CliShe07}.  But they just come down to an
old-fashioned (and usually violent) conflict with observation.  This
has already forced us to abandon or modify (and complicate) some of
the simplest approaches.

\paragraph{Anthropic selection effects}
Vacua without observers will not contribute to the statistical
ensemble that determines what observations are likely.  For example,
most vacua in the landscape have a cosmological constant of order
unity.  They will be about one Planck length in size and contain at
most a handful of quantum states~\cite{RMP}.  Even without strong
assumptions about what observers look like, we can be quite sure that
these vacua will not be observed.

Quantifying this selection effect is a challenging task, for two
reasons.  First, it inevitably leads to the problematic question of
what constitutes an observer.  Supposing we can agree on some
criterion, we can then ask what a typical observer sees.  This
involves tallying observers, or observations, across the whole
universe.  Each vacuum bubble is an infinite open universe in itself,
so it contains either zero observers or infinitely many.  To define
the relative abundance of different observations, a cutoff is
required.  This entangles us, once more, in the measure problem
discussed earlier.

One possibility is to use one measure to compute only the abundance of
vacuum bubbles, and a separate regularization scheme inside each
bubble to define the abundance of observers.  An example of the latter
is the number of observers per baryon, or per
photon~\cite{Efs95,Vil95,Wei96,MarSha97,Vil04,Wei05}.  But this
quantity seems somewhat arbitrary, and it may not be well-defined
throughout the string landscape, since not all vacua will contain the
specified reference particles.  Another strategy is to calibrate a
``unit comoving volume''~\cite{GarSch05}, but it is not clear that
this can be rigorously defined.

The problem simplifies if we restrict to vacua that differ from ours
only in $\rho_\Lambda$.  Then all of the above schemes are equivalent.
Assuming only that observers require galaxies, they prefer a
cosmological constant about three orders of magnitude larger than the
observed value.  The agreement improves if more detailed assumptions
about observers are made.  It worsens when additional parameters, in
particular the primordial density contrast, are allowed to vary (as
seems inevitable in the string
landscape)~\cite{TegRee97,BanDin03,GarVil05}.

Another possibility is to use the same measure for counting both
vacuum bubbles and the observers in them.  For example, the number of
observers in a single causally connected region is already finite in
any vacuum with nonzero cosmological constant, by Eq.~(\ref{eq-rlam}).
(In the string landscape, vacua with $\rho_\Lambda=0$ have unbroken
supersymmetry.  This would seem to preclude any form of condensed
matter, and thus, presumably, observers.)

The causal diamond measure has another, quantitative advantage: the
size of the causally connected region depends on the cosmological
constant through Eq.~(\ref{eq-trlam}).  The smaller the cosmological
constant, the larger the causally connected region, and the more
complexity it allows.  Thus, small values of $\rho_\Lambda $ are not
just enforced by galaxy formation.  They are favored more generally
because they allow more room, and more time, for observers.

Let us again restrict to the set of vacua identical to ours except for
$\rho_\Lambda$, and ask what probability distribution for
$\rho_\Lambda $ the causal diamond measure predicts.  One finds that
the favored value of $\rho_\Lambda$ is that in which vacuum energy
comes to dominate the energy density around the time when most
observations are made.  This result generalizes to very different
vacua, so the coincidence problem (Sec.~\ref{sec-coincidence}) is
actually the primary problem solved by this measure.

The numerical value of $\rho_\Lambda$ then depends on the low energy
physics determining the epoch of observers.  With values larger than
of order $10^{-123}$, galaxies are expelled from the horizon before
stars and observers form; smaller values are simply less frequent in
the landscape.  Thus, the causal diamond predicts a probability
distribution for $\rho_\Lambda$ that is in remarkable agreement with
the observed value~\cite{BouHar07}.

One would not expect this agreement to worsen when the primordial
density contrast is allowed to vary as well.  Larger density
perturbations speed up galaxy formation, undermining the necessity for
small $\rho_\Lambda$.  But in the causal diamond measure, galaxy
formation (while still a necessary condition for observers) is no
longer the dominant constraint on $\rho_\Lambda$.

The problem of characterizing observers, especially in vacua very
different from ours, remains challenging.  A surprisingly good
approximation, in examples studied so far, is to replace the number of
observers by the amount of entropy that is produced by a given vacuum
in the causal diamond, $\Delta S_{\rm CD}$.  Whatever observers are,
they must obey the laws of thermodynamics.  As they compute, store and
retrieve information, they convert free energy into entropy.  Of
course, not all entropy is produced by observers.  But $\Delta S_{\rm
  CD}$ can be thought of as an upper bound on the complexity of a
spacetime region. At least on average, it might be proportional to the
number of observations made.

In vacua similar to ours, where we do have ideas about what observers
look like, this {\em Causal Entropic Principle\/} agrees with
conventional weighting by the observers present in the causal
diamond~\cite{BouHar07}.  Remarkably, this allows us to explain
$\rho_\Lambda$ without even assuming that observers require galaxies.
One finds that $\Delta S_{\rm CD}$ is dominated by infrared radiation
from dust heated by starlight.  Much less entropy would have been
produced in the absence of heavy elements, or stars, or galaxies.
Thus, we may have identified a primitive and universal criterion
capable of capturing conditions often assumed (by hand) as necessary
for complex life.  The causal diamond and $\Delta S_{\rm CD}$ are
well-defined in any vacuum, and it may be possible to estimate them,
at least on average, even in distant regions of the landscape.

Tools like this will be crucial if we want to progress from
conditional predictions, which correlate various features of our own
vacuum, to a fundamental understanding of their origin.  In the string
landscape, a scale like $10^{-123}$ ultimately must arise from the
density of its discretuum and the range of complexity of its particle
physics.

\acknowledgments I would like to thank M.~Aganagic, R.~Bean,
B.~Freivogel, J.~Polchinski, M.~Porrati, and L.~Susskind for
discussions and correspondence.

\bibliographystyle{board}
\bibliography{all}
\end{document}